\documentclass[sigconf,nonacm]{acmart}
\usepackage{amsmath,mathtools}
\usepackage{booktabs}
\usepackage{colortbl}
\usepackage{graphicx}
\usepackage{subcaption}
\usepackage{enumitem}
\usepackage{multirow}
\usepackage{placeins}
\usepackage{tabularx}
\usepackage{tikz}
\usetikzlibrary{arrows.meta,positioning}
\usepackage[nameinlink,capitalise]{cleveref}
\usepackage{xspace}

\newcommand{\sid}{\mathrm{SID}}
\newcommand{\history}{\mathcal{H}}

\newcommand{\loss}{\mathcal{L}}
\newcommand{\method}{M\&B\xspace}

\newcommand{\cmark}{\textcolor{green!50!black}{\ensuremath{\boldsymbol{\checkmark}}}}
\newcommand{\result}[2]{#1{\scriptsize\color{black!45}\ensuremath{\,\pm\,#2}}}

\newcolumntype{Y}{>{\raggedright\arraybackslash}X}
\setlist{nosep,leftmargin=*}

\makeatletter
\g@addto@macro\UrlBigBreaks{\do\@}
\makeatother

\title[Merge and Branch: Efficient Tree-Structured Supervised Fine-Tuning of LLMs for Recommendation]{\texorpdfstring{Merge and Branch: Efficient Tree-Structured\\ Supervised Fine-Tuning of LLMs for Recommendation}{Merge and Branch: Efficient Tree-Structured Supervised Fine-Tuning of LLMs for Recommendation}}
\author{Yunqin Zhu}
\email{yzhu812@gatech.edu}
\affiliation{
  \institution{Georgia Institute of Technology}
  \city{Atlanta}
  \state{GA}
  \country{USA}
}
\author{Yao Xie}
\email{yao.xie@isye.gatech.edu}
\affiliation{
  \institution{Georgia Institute of Technology}
  \city{Atlanta}
  \state{GA}
  \country{USA}
}
\hypersetup{
  pdfauthor={Yunqin Zhu and Yao Xie},
  pdftitle={Merge and Branch: Efficient Tree-Structured Supervised Fine-Tuning of LLMs for Recommendation},
  pdfsubject={},
  pdfkeywords={},
  pdfcreator={LaTeX},
  pdfproducer={pdfTeX}
}

\begin{document}
\begin{abstract}
LLM-based generative recommendation reads long, text-rich interaction histories but generates only short semantic identifiers (SIDs). This asymmetry makes supervised fine-tuning inefficient: encoding the history is expensive, yet each history usually provides supervision for only one target. We present Merge and Branch (\method) to reduce input cost and learn more from each encoded history. \emph{Merge} compresses each historical item's SID and text into a small number of learned tokens. \emph{Branch} uses the compressed history to predict multiple target SIDs. To share history computation while keeping these targets independent, \method organizes each example as a shallow, wide tree. We develop tree FlashAttention to execute this layout efficiently, processing the shared history once and propagating gradients from every branch for end-to-end training. A tree-structured key/value cache extends history sharing to constrained decoding. Experiments on public benchmarks show that \method outperforms strong recommendation baselines while substantially improving training and decoding efficiency.
\end{abstract}

\maketitle
\raggedbottom

\section{Introduction}
\label{sec:introduction}

\begin{figure}[t]
  \centering
  \includegraphics[width=\columnwidth]{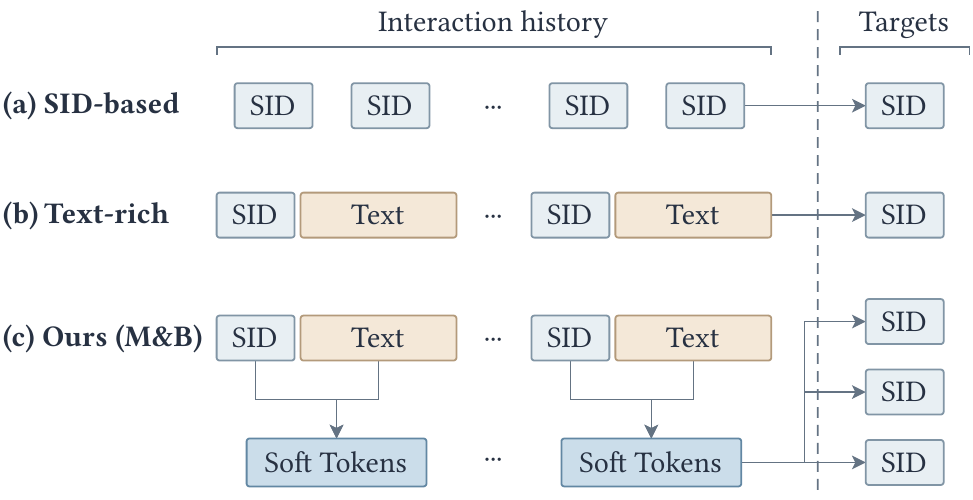}
  \caption{Generative recommendation paradigms. M\&B pairs compact SID--text inputs with independent SID targets.}
  \label{fig:paradigms}
  \Description{Three rows compare SID history to one SID target, SID and text history to one SID target, and merged SID and text tokens to parallel SID targets.}
\end{figure}

Generative recommendation retrieves catalog items by predicting their semantic identifiers (SIDs) from a user's interaction history~\citep{rajput2023tiger}. SIDs provide compact item representations, while item text, such as titles and categories, supplies additional content that large language models (LLMs) can use to interpret user preferences~\citep{geng2022p5,zheng2024lcrec,hong2025eagerllm}. Incorporating this text, however, creates an imbalance between input cost and supervision: a model may process thousands of history tokens to predict a next-item SID of only a few tokens. This raises a practical question: how can we reduce the cost of processing each history while learning more from it?

Reducing input cost calls for a compact representation that retains SID and textual information as well as the sequence of interactions. Learning more from each history also requires looking beyond the next item, which provides only one observation of a user's preferences. Additional future interactions in the training sequence offer further observed targets. A trained teacher recommender can also propose synthetic targets, transferring collaborative information through its item predictions. These additional targets should each be predicted from the observed history, without requiring other target items at inference. Naively training a separate example for every target satisfies this requirement but repeatedly processes the same history. We therefore seek a training structure that shares the history across predictions while keeping the targets separate.

To address the challenges, we propose \textbf{Merge and Branch (\method)}, which combines a compact history with multiple independent target branches (Figure~\ref{fig:paradigms}). \emph{Merge} compresses each historical item's SID and text into a small number of learned tokens. SID-conditioned queries select content within each item, and the resulting tokens remain in chronological order so that the LLM can model dependencies across interactions. \emph{Branch} uses this compact history to predict the next item, additional observed future items, and teacher-proposed synthetic targets. Each target attends to the shared history and its own preceding SID tokens, forming a shallow tree in which targets cannot read one another. Losses from all branches jointly train the history representation and the LLM.

Efficient execution is essential to realizing the benefit of this structure. We develop \emph{tree FlashAttention} (tree FA) with forward and backward kernels that support the branch visibility pattern without materializing a dense attention matrix. The kernels compute the shared history once and accumulate gradient contributions from every branch. The same sharing principle extends to constrained beam search. Beam hypotheses use the same history but differ in their generated SIDs, so our tree-structured key/value (KV) cache stores the history once and maintains separate states for each hypothesis.

Experiments on three Amazon domains show that \method outperforms the sequential and generative recommendation baselines in our evaluation. Ablations show that learning from multiple targets improves recommendation quality with compact history representations. Efficiency measurements demonstrate substantially lower training cost than processing uncompressed SID and text histories. Shared-prefix caching also reduces beam-search latency and memory usage, with the largest benefits at long contexts.

Our contributions are threefold:
\nopagebreak
\begin{itemize}
  \item We introduce \method, an LLM-based generative recommendation framework that reduces history input cost and learns from multiple targets through a shared history.
  \item We design an interaction token merger and a multi-target training objective that combines observed and synthetic targets while keeping target predictions independent.
  \item We develop tree FlashAttention with forward and backward support and a shared-prefix KV cache to execute multi-target training and constrained beam search efficiently.
\end{itemize}

\section{Related Work}
\label{sec:related-work}

\paragraph{Generative recommendation.}
Generative recommendation depends on both item representation and the choice of prediction targets. TIGER quantizes item text embeddings into hierarchical SIDs and generates the next item's identifier \citep{rajput2023tiger}. LETTER and ETEGRec incorporate collaborative information into item tokenization, while EAGER jointly models behavioral and semantic identifiers \citep{wang2024letter,liu2025etegrec,wang2024eager}. ActionPiece learns context-dependent tokenizations of action sequences \citep{hou2025actionpiece}. HSTU scales autoregressive sequence modeling for recommendation, and OneRec generates item sequences at the session level \citep{zhai2024hstu,deng2025onerec}. Beyond next-item supervision, relevance-aware training learns from several future items, and UniSGR predicts multiple SIDs alongside behavioral objectives \citep{bacciu2023relevance,sun2026unisgr}. Alternatives to autoregressive generation include parallel SID prediction~\citep{hou2025rpg} and diffusion-based recommendation, ranging from interaction reconstruction~\citep{wang2023diffrec,zhu2024giffcf} through continuous item representation generation~\citep{yang2023dreamrec,li2023diffurec} to masked discrete SID generation~\citep{liu2026diffgrm}. \method retains the SID output space and focuses on learning from multiple independent targets over a compact history.

\paragraph{LLM recommenders.}
LLM recommenders differ in how they incorporate item information and construct supervision. P5 and P5-CID formulate recommendation tasks with T5-family encoder--decoder models \citep{geng2022p5,hua2023index}, whereas TALLRec, LLaRA, and CoLLM adapt decoder-only LLMs for candidate-conditioned prediction or selection \citep{bao2023tallrec,liao2024llara,zhang2025collm}. For retrieval from the full catalog, BIGRec grounds generated text in catalog items, LC-Rec and EAGER-LLM generate discrete identifiers, and SETRec generates sets of continuous item embeddings \citep{bao2025bigrec,zheng2024lcrec,hong2025eagerllm,lin2025setrec}. GRAM incorporates collaborative information into item descriptions and fuses separately encoded prompts during decoding \citep{lee2025gram}. Richer supervision is also studied in language modeling through multi-token prediction \citep{gloeckle2024multitoken}. GenRec addresses both input cost and supervision: it linearly merges historical SID embeddings and serializes target items into a page \citep{zou2026genrec}. \method instead compresses SID and text jointly through learned attention and keeps each target independent of the others. All target losses train the merger and LLM through one shared history.

\paragraph{Tree attention and KV caching.}
Prefix reuse reduces repeated computation in both inference and training. SGLang and ChunkAttention share KV states across requests with common prefixes \citep{zheng2024sglang,ye2024chunkattention}. SpecInfer and Medusa use tree masks to verify speculative tokens, while DeFT develops IO-aware attention for tree-structured decoding \citep{miao2024specinfer,cai2024medusa,yao2025deft}. In recommendation, AtSpeed verifies speculative beam candidates with a tree mask, and UniSGR's STARK reorganizes KV states for SID beam search \citep{lin2025atspeed,sun2026unisgr}. Training additionally requires gradients from every continuation to reach the shared prefix. Tree Training packs branching trajectories for shared forward and backward computation, while AReaL-DTA traverses rollout trees during policy training \citep{wang2025treetraining,zhang2026arealdta}. \method targets the shallow trees formed by long recommendation histories and short SID branches. Building on FlashAttention \citep{dao2022flashattention}, its tree kernels support multi-target fine-tuning, and its KV cache applies the same visibility structure to constrained beam search. Table~\ref{tab:tree-capabilities} \mbox{compares these capabilities}.

\begin{table}[t]
  \centering
  \renewcommand{\arraystretch}{1.08}
  \setlength{\tabcolsep}{3pt}
  \caption{Support for tree-structured training and decoding. Fwd./Bwd.: FA-style forward/backward kernels, excluding dense masks and path traversal approaches.}
  \label{tab:tree-capabilities}
  \begin{tabularx}{\columnwidth}{@{}l*{5}{>{\centering\arraybackslash}X}@{}}
    \toprule
    Method & Fwd. & Bwd. & Cache & SFT & Beam \\
    \midrule
    SGLang~\citep{zheng2024sglang} & & & \cmark & & \\
    DeFT~\citep{yao2025deft} & \cmark & & \cmark & & \\
    Tree Training~\citep{wang2025treetraining} & \cmark & \cmark & & \cmark & \\
    AtSpeed~\citep{lin2025atspeed} & & & & & \cmark \\
    UniSGR~\citep{sun2026unisgr} & & & \cmark & \cmark & \cmark \\
    \midrule
    \textbf{\method} & \cmark & \cmark & \cmark & \cmark & \cmark \\
    \bottomrule
  \end{tabularx}
\end{table}

\section{Merge and Branch}
\label{sec:method}

\begin{figure*}[t]
  \centering
  \includegraphics[width=0.9\textwidth]{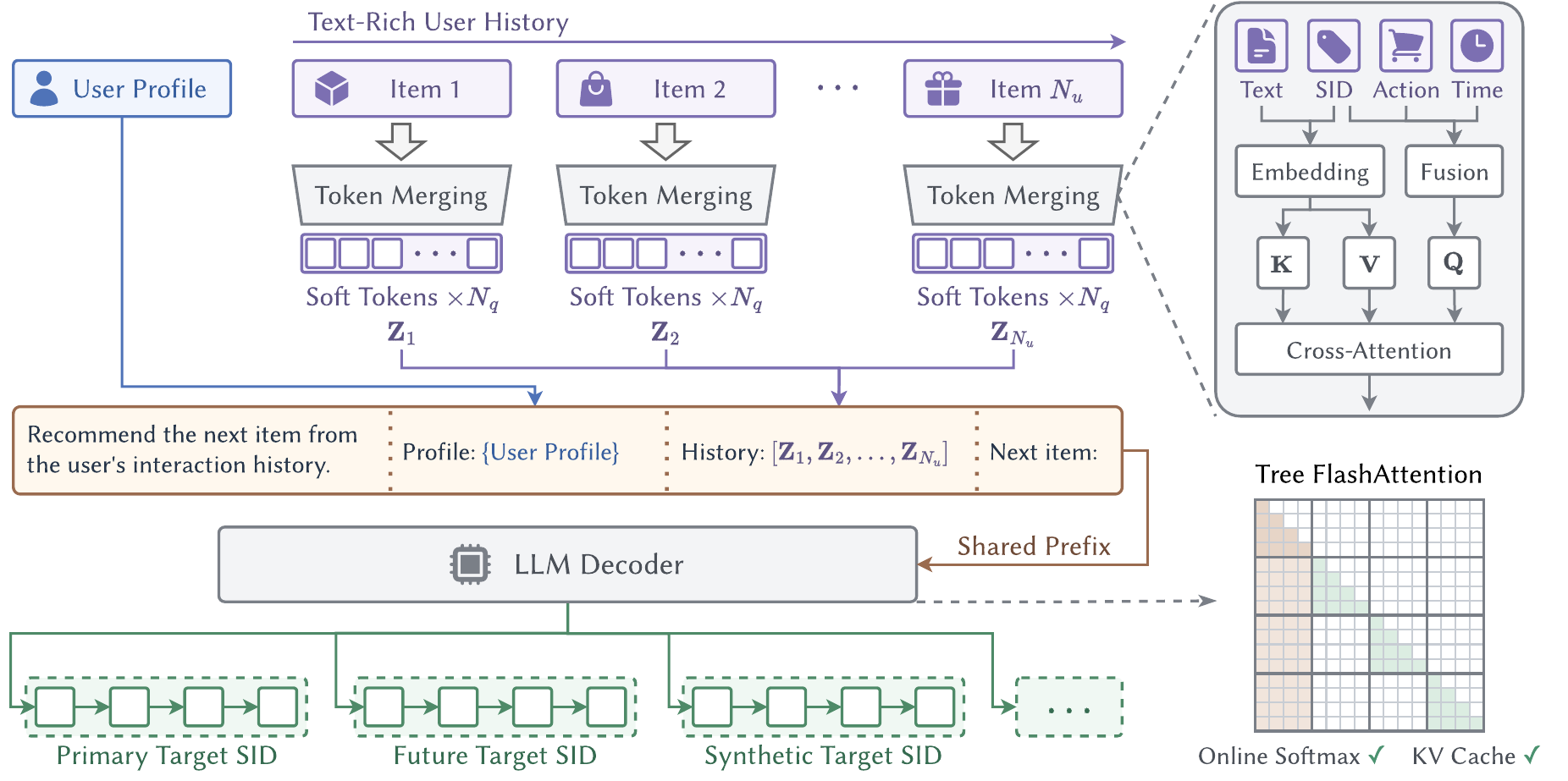}
  \caption{Overview of \method. A lightweight cross-attention merges each interaction into $N_q$ soft tokens, which join the instruction and optional user profile to form the LLM prompt. Tree FlashAttention shares this prefix across independent primary, future, and synthetic SID branches during training, while KV caching reuses the prefix during beam search.}
  \Description{A text-rich interaction history is compressed item by item into ordered soft tokens. An inset shows SID, action, and time information conditioning queries that attend to SID and text keys and values. The soft tokens join the task instruction and optional user profile in a shared LLM prefix. Primary, future, and synthetic SID branches read this prefix independently. The tree attention matrix shows causal attention within the prefix and within each branch, with no attention between sibling branches. Online softmax supports attention computation, and KV caching shares the prefix during decoding.}
  \label{fig:overview}
\end{figure*}

\subsection{Problem Formulation and Overview}

For a user $u$, let $\history_u=(v_{u,1},\ldots,v_{u,N_u})$ denote the observed item history, accompanied by item text and any available interaction metadata. Each catalog item $y$ has a fixed-length semantic identifier $\sid(y)=(s_1(y),\ldots,s_L(y))$. The model recommends an item by generating its identifier autoregressively from the observed history.

M\&B addresses long inputs through interaction token merging and limited supervision through multi-target training (Figure~\ref{fig:overview}). The merger maps each historical item's SID and text embeddings to $N_q$ soft tokens, forming a compact prefix $p_u$ of length $T_u$. For multi-target training, we attach one SID sequence per target to this prefix. A pretrained LLM processes the prefix and target branches with tree attention at every layer, so each target can read the history without reading other targets. The merger and LLM learn jointly from these predictions. At inference, the same structure lets beam hypotheses share the history's KV cache.

\subsection{Interaction Token Merging}
\label{sec:merge}

Item text supplies useful content but can dominate the history's token count. We compress each interaction independently to shorten the history while keeping the sequence of items explicit. A single merger, shared across items, maps the $M_i$ valid SID and text embeddings $\mathbf X_i\in\mathbb R^{M_i\times d}$ to $N_q$ soft tokens $\mathbf Z_i\in\mathbb R^{N_q\times d}$, where $d$ is the LLM hidden size. The outputs remain in chronological order: the merger selects information within each item, and the LLM models dependencies across items.

\paragraph{SID anchor.}
An item's SID identifies it in the catalog, while its text describes its content. We first combine the SID embeddings into an anchor for the compressed representation:
\begin{equation}
  \begin{aligned}
  \bar{\mathbf e}_i
    &=\frac{1}{L}\sum_{\ell=1}^{L}\mathbf e_{i,\ell}, \\
  \mathbf g_i
    &=\operatorname{Concat}_{\ell=1}^{L}
      \mathcal N_{\mathrm{SID}}(\mathbf e_{i,\ell}), \\
  \mathbf c_i
    &=\bar{\mathbf e}_i+
      \mathbf W_2\operatorname{SiLU}(\mathbf W_1\mathbf g_i).
  \end{aligned}
  \label{eq:merge-anchor}
\end{equation}
Here $\mathbf e_{i,\ell}=E(s_\ell(v_{u,i}))$ is the SID embedding at level $\ell$, $E$ is the LLM token embedding table, and $\mathcal N_{\mathrm{SID}}$ is RMSNorm. The mean gives a direct residual path, while concatenation lets the MLP distinguish SID levels and learn how they relate. When interaction metadata is available, the fusion step in Figure~\ref{fig:overview} adds learned action-type and time-bucket embeddings to $\mathbf c_i$ before the following operations.

\paragraph{Content selection.}
We use cross-attention to select content from the item's SID and text tokens. Each shared learned offset $\mathbf q_r$ combines with the SID anchor to form a query $\widetilde{\mathbf q}_{i,r}$. Queries and keys have dimension $d_a$, and values have dimension $d_v$. Field embeddings distinguish SID and text fields, while sinusoidal encodings indicate token positions. Both are added only to the keys, guiding selection without changing the values. With field label $f_{i,m}$ for token $m$ and separate RMSNorm layers $\mathcal N_q$ and $\mathcal N_{kv}$, we compute
\begin{equation}
  \begin{aligned}
  \widetilde{\mathbf q}_{i,r}
    &=\mathbf W_q\mathcal N_q(\mathbf c_i+\mathbf q_r), \\
  \widetilde{\mathbf k}_{i,m}
    &=\mathbf W_k\mathcal N_{kv}(\mathbf x_{i,m})
      +\mathbf e^{\mathrm{field}}_{f_{i,m}}
      +\mathbf W_{\mathrm{pos}}\operatorname{PE}(m), \\
  \widetilde{\mathbf v}_{i,m}
    &=\mathbf W_v\mathcal N_{kv}(\mathbf x_{i,m}), \\
  \alpha_{i,r,m}
    &=\operatorname{softmax}_{m}\!\left(
      \frac{\widetilde{\mathbf q}_{i,r}^{\top}
      \widetilde{\mathbf k}_{i,m}}{\sqrt{d_a}}\right).
  \end{aligned}
  \label{eq:merge-qkv}
\end{equation}
Because each query depends on the item's SID anchor, it can select different content across items despite sharing the learned offsets.

\paragraph{Residual output.}
We project the selected content with $\mathbf W_o\in\mathbb R^{d\times d_v}$ and add it to the anchor in the LLM hidden space:
\begin{equation}
  \mathbf z_{i,r}
  =\mathbf c_i+\mathbf W_o
    \sum_{m=1}^{M_i}\alpha_{i,r,m}\widetilde{\mathbf v}_{i,m}.
  \label{eq:merge-output}
\end{equation}
This residual retains the anchor; learned offsets guide content selection but are not added to the output. We initialize $\mathbf W_2$ and $\mathbf W_o$ to zero, so training starts from the mean SID embedding plus any action-type and time-bucket embeddings and learns the corrections. Increasing $N_q$ yields more summaries with the same merger.

We replace each item's original fields in the prompt with $\mathbf Z_i=[\mathbf z_{i,1},\ldots,\mathbf z_{i,N_q}]^{\top}$, retaining item boundaries, the task instruction, and any user profile. Compared to Q-Former, which starts from shared learned query embeddings~\citep{li2023blip2}, our merger constructs its queries from the item's SID anchor and learned offsets. LONGER merges groups of adjacent interactions~\citep{chai2025longer}, while GenRec and Token Factory map item features to soft tokens~\citep{zou2026genrec,chen2026tokenfactory}. Our merger uses SID-conditioned queries to compress SID and text within each interaction. The LLM can thus model the item sequence without reading the full text.

\subsection{Multi-Target Training}
\label{sec:branch}
\label{sec:multi-target-supervision}

\paragraph{Supervision sources.}
A single next-item target captures only one observed preference for a history. Alongside the primary next item $y_u^{(0)}$, we use additional future targets $\mathcal F_u$ from the training sequence and synthetic targets $\mathcal S_u$ from an established recommendation model, e.g., SASRec \cite{kang2018sasrec}. The former capture further observed preferences; the latter transfer the teacher's recommendations in a knowledge distillation flavor. The teacher's checkpoint can be selected by validation ranking performance and kept frozen during training. We retain observed target occurrences, including repeated items, and add synthetic targets not already among them, giving $K_u=1+|\mathcal F_u|+|\mathcal S_u|$ branches.

\paragraph{Training objective.}
For each target, the model predicts the SID from the shared prefix and that target's preceding SID tokens:
\begin{equation}
  \ell_u(y)=-\frac{1}{L}\sum_{\ell=1}^{L}
  \log p_\theta\!\left(s_\ell(y)\mid p_u,s_{<\ell}(y)\right).
  \label{eq:sid-likelihood}
\end{equation}
The softmax at level $\ell$ is over that level's SID codebook. The number of auxiliary targets can vary, but it should not determine their total influence on training. We therefore assign weights $\lambda_{\mathrm F}$ and $\lambda_{\mathrm S}$ to the two groups and distribute each weight uniformly over its available targets. For a nonempty group $\mathcal G$, let $\bar\ell_u(\mathcal G)$ denote its mean loss:
\begin{equation}
  \begin{aligned}
    \bar\ell_u(\mathcal G)
      &=\frac{1}{|\mathcal G|}\sum_{y\in\mathcal G}\ell_u(y), \\[2pt]
    \loss_{\mathrm{SID}}(u)
      &=\frac{\ell_u(y_u^{(0)})
        +\lambda_{\mathrm F}\bar\ell_u(\mathcal F_u)
        +\lambda_{\mathrm S}\bar\ell_u(\mathcal S_u)}{Z_u}.
  \end{aligned}
  \label{eq:branch-objective}
\end{equation}
Empty groups contribute no term, and $Z_u$ sums the active group weights, including the primary weight of one. Minimizing the mean loss over histories trains the merger, token embeddings, and LLM jointly. The weights thus control each source's supervision strength separately from its target count.

\paragraph{Tree-structured training.}
Each target should be predicted from the observed history, not from other targets that are unavailable at inference. Processing targets separately satisfies this requirement but repeats the history computation; concatenating them with causal attention introduces dependencies between targets. We instead attach one teacher-forced SID branch per target to $p_u$. Tree-structured training lets each branch read the entire prefix and its own preceding SID tokens, but not another branch. All branches share the LLM parameters and output projection.

Sharing computation also requires consistent positions and prediction locations. Each branch starts immediately after the history in positional space: its input tokens receive zero-based positions $T_u,\ldots,T_u+L-1$, regardless of their location in the packed sequence. The prefix's last hidden state predicts the first SID token of every target using one shared distribution. Within a branch, the state of $s_\ell(y)$ predicts $s_{\ell+1}(y)$. Each target is thus treated as a continuation of the same history. We project only supervised states onto their SID codebooks and skip projection for other history states.

\subsection{Tree FlashAttention}
\label{sec:tree-attention}

A dense implementation of tree attention would still store a quadratic attention matrix. Tree FlashAttention (tree FA) applies the tree structure through tiled attention and online softmax~\citep{dao2022flashattention}.

The packed sequence contains $T_u+K_uL$ positions: one prefix followed by contiguous SID branches. Prefix tokens use causal attention; branch tokens see the prefix and the causal part of their own branch. For a branch token at packed index $a$, let $\beta(a)$ identify its branch and $\delta(a)$ its offset. The allowed key indices are
\begin{equation}
  \mathcal V_u(a)=
  \begin{cases}
    \{0,\ldots,a\}, & a<T_u,\\
    \{0,\ldots,T_u-1\}\cup\mathcal B_u(a), & a\ge T_u,
  \end{cases}
  \label{eq:tree-visibility}
\end{equation}
where $\mathcal B_u(a)=\{b\ge T_u:\beta(b)=\beta(a),\ \delta(b)\le\delta(a)\}$ contains the causal part of that branch. Applying this rule at every layer prevents targets from affecting the prefix or one another. With matching position indices, each branch sees the same inputs as a separately executed continuation. Tokenwise projections, normalization, and feed-forward layers preserve this correspondence.

\paragraph{Forward pass.}
Prefix lengths, branch counts, and SID depth are enough to describe the layout, so the kernel does not need a dense mask. It evaluates the visibility rule within each tile, excluding sibling branches and padding before the softmax update. Each branch query normalizes jointly over the prefix and its own SID keys, preserving ordinary attention over the visible tokens. Running softmax statistics combine successive tiles without storing the attention matrix; the output and log-sum-exp are saved for backward. The kernel supports grouped-query attention and varying prefix lengths and branch counts across examples.

\paragraph{Backward pass.}
Because every branch uses the same prefix, its loss must contribute to the prefix's gradient. At one attention layer, partition the attention probabilities $\mathbf A$ into prefix rows $p$ and branch rows $j$. If $\mathbf G_p$ and $\mathbf G_j$ are the corresponding output gradients, including the loss weights, the gradient of prefix values is
\begin{equation}
  \nabla_{\mathbf V_p}\loss
    =\mathbf A_{pp}^{\top}\mathbf G_p
     +\sum_{j=1}^{K_u}\mathbf A_{jp}^{\top}\mathbf G_j.
  \label{eq:tree-value-gradient}
\end{equation}
Prefix keys likewise receive contributions from prefix and branch queries through the softmax derivative. Tree FA reconstructs the probabilities tile by tile from the saved log-sum-exp and sums these contributions into the shared key/value gradients. Together with the query and residual paths, these gradients reach the merger through the shared prefix computation. Every target therefore trains the history representation while reusing its computation. Appendix~\ref{sec:method-details} gives the online-softmax recurrence and full gradients.

\paragraph{Computational complexity.}
The main saving comes from computing the history once rather than once per target. For $K$ targets of SID depth $L$ and a prefix of length $T$, independent execution processes $K(T+L)$ token positions; the tree processes $T+KL$. This shares prefix projections, feed-forward computation, and saved activations. The number of visible query--key pairs changes from
\begin{equation}
  \begin{aligned}
  C_{\mathrm{independent}}
    &=K\!\left[\frac{T(T+1)}{2}+LT+\frac{L(L+1)}{2}\right], \\[2pt]
  C_{\mathrm{tree}}
    &=\frac{T(T+1)}{2}+KLT+\frac{KL(L+1)}{2}.
  \end{aligned}
  \label{eq:tree-cost}
\end{equation}
Only attention within the history is reused; each branch still reads the history. These counts describe visible pairs rather than hardware FLOPs, since tiles may include masked entries. In summary, token merging shortens the prefix, tree-structured training shares its computation across targets, and tree FA executes this layout without materializing a dense attention matrix.

\subsection{Shared-Prefix Beam Search}
\label{sec:tree-kv-cache}

Beam hypotheses differ in their generated SID tokens but share the same observed history. Duplicating that history's KV states for every beam is therefore unnecessary. We prefill $p_u$ once and retain its keys and values at every LLM layer. A tree-structured KV cache stores this prefix once and keeps separate SID states for each hypothesis. As in training, every active query reads the prefix and its own SID history under the same visibility and position rules.

Not every SID sequence identifies a catalog item, so a GPU-resident trie restricts each expansion to valid continuations. Beam scores accumulate log-probabilities normalized over the full level codebook; the trie excludes invalid extensions without renormalizing over its children. After selecting the surviving beams, the decoder gathers their SID states from the selected parents. Only these short branch states are copied; the shared prefix stays in place. All beams use the same LLM and level-specific SID projection.

Let $W$ denote beam capacity, distinct from the number of training targets $K$. The cache holds at most $T+WL$ token states per layer, rather than $W(T+L)$ with duplicated histories, saving $(W-1)T$ stored prefix states. Each hypothesis still computes attention to the history, but wider beams no longer require more copies of it. Training and decoding thus reuse the same tree structure: training combines gradients through a common history, while decoding retains that history as hypotheses change.

\section{Experiments}
\label{sec:experiments}

We organize the experiments around four research questions:
\begin{itemize}
  \item \textbf{RQ1:} How does M\&B compare with existing methods?
  \item \textbf{RQ2:} How do token merging and auxiliary supervision affect recommendation accuracy?
  \item \textbf{RQ3:} How efficient are training and decoding with M\&B?
  \item \textbf{RQ4:} How do backbone choice and pretraining settings affect convergence?
\end{itemize}

\subsection{Public Benchmarks (RQ1)}

\paragraph{Datasets.}
Following TIGER~\citep{rajput2023tiger}, we use three domains from the Amazon 2014 5-core reviews~\citep{he2016ups}: Beauty, Sports \& Outdoors (Sports), and Toys \& Games (Toys). We order each user's interactions chronologically. Table~\ref{tab:dataset-statistics} summarizes the datasets.

\begin{table}[t]
  \centering
  \renewcommand{\arraystretch}{1.06}
  \setlength{\tabcolsep}{3.5pt}
  \caption{Dataset statistics. Len.: mean interactions per user; Density: observed fraction of user--item pairs.}
  \label{tab:dataset-statistics}
  \begin{tabular*}{\columnwidth}{@{\extracolsep{\fill}}lrrrrr@{}}
    \toprule
    Dataset & \# Users & \# Items & \# Inters. & Len. & Density \\
    \midrule
    Beauty & 22,363 & 12,101 & 198,502 & 8.9 & 0.0734\% \\
    Sports & 35,598 & 18,357 & 296,337 & 8.3 & 0.0453\% \\
    Toys & 19,412 & 11,924 & 167,597 & 8.6 & 0.0724\% \\
    \bottomrule
  \end{tabular*}
\end{table}

\paragraph{Metrics and protocol.}
We report Recall and NDCG at cutoffs 5 and 10 over the full item catalog. Each test instance has one relevant item, so Recall@$K$ equals Hit Rate@$K$; invalid or unresolved predictions count as misses. Following TIGER~\citep{rajput2023tiger}, we reserve each user's last interaction for testing and the penultimate interaction for validation, using the rest for training. Validation NDCG@10 selects the maximum history length from $\{20,50\}$ and each M\&B run's checkpoint. The main benchmark uses the best of five M\&B seeds by validation NDCG@10 for all four test metrics.

\paragraph{Implementation.}
We fine-tune Qwen3-1.7B jointly with the token merger. We quantize Sentence-T5-XXL item embeddings with a three-level RQ-VAE using 256 codes per level and append a deterministic collision suffix, yielding four-token SIDs. Item text comprises the title, category, and description. The main benchmark uses one merged token per historical item and beam width 50. Each training example supervises the next item, at most one additional future item, and two synthetic targets proposed by a frozen SASRec teacher trained with cross-entropy and selected on validation data. The future and synthetic groups receive weights $\lambda_{\mathrm F}=0.25$ and $\lambda_{\mathrm S}=1.0$. We train for four epochs in BF16 with AdamW and batch size 128, using learning rate $10^{-4}$ for the backbone and merger and $10^{-3}$ for the tied token embeddings. See Appendix~\ref{sec:public-configuration} for details.

\begin{table*}[t]
  \centering
  \setlength{\tabcolsep}{2.5pt}
  \renewcommand{\arraystretch}{1.06}
  \caption{Next-item recommendation on Amazon. R/N: Recall/NDCG. Underline: best baseline; bold: best overall. \textit{Improv.}: relative gain over the strongest baseline.}
  \label{tab:public-results}
  \begin{tabularx}{\textwidth}{@{}ll*{12}{>{\centering\arraybackslash}X}}
    \toprule
    \multirow{2}{*}{Family} & \multirow{2}{*}{Method}
      & \multicolumn{4}{c}{Beauty}
      & \multicolumn{4}{c}{Sports}
      & \multicolumn{4}{c}{Toys} \\
    \cmidrule(lr){3-6}\cmidrule(lr){7-10}\cmidrule(lr){11-14}
      & & R@5 & N@5 & R@10 & N@10
      & R@5 & N@5 & R@10 & N@10
      & R@5 & N@5 & R@10 & N@10 \\
    \midrule
    \multirow{3}{*}{ID-based}
      & SASRec
      & 0.0537 & 0.0376 & 0.0788 & 0.0456
      & 0.0293 & 0.0198 & 0.0432 & 0.0243
      & 0.0661 & 0.0479 & 0.0860 & 0.0543 \\
      & SASRec-CE
      & \underline{0.0680} & \underline{0.0496} & \underline{0.0932} & \underline{0.0577}
      & \underline{0.0375} & \underline{0.0269} & 0.0537 & \underline{0.0322}
      & \underline{0.0745} & \underline{0.0554} & \underline{0.0988} & \underline{0.0633} \\
      & BERT4Rec
      & 0.0372 & 0.0260 & 0.0554 & 0.0319
      & 0.0161 & 0.0106 & 0.0251 & 0.0135
      & 0.0338 & 0.0242 & 0.0475 & 0.0286 \\
    \midrule
    \multirow{4}{*}{SID-based}
      & TIGER
      & 0.0454 & 0.0321 & 0.0648 & 0.0384
      & 0.0264 & 0.0181 & 0.0400 & 0.0225
      & 0.0521 & 0.0371 & 0.0712 & 0.0432 \\
      & LETTER
      & 0.0431 & 0.0286 & 0.0672 & 0.0364
      & 0.0275 & 0.0180 & 0.0428 & 0.0229
      & 0.0436 & 0.0294 & 0.0679 & 0.0372 \\
      & EAGER
      & 0.0618 & 0.0451 & 0.0836 & 0.0525
      & 0.0281 & 0.0184 & 0.0441 & 0.0236
      & 0.0584 & 0.0464 & 0.0714 & 0.0505 \\
      & ETEGRec
      & 0.0511 & 0.0342 & 0.0785 & 0.0430
      & 0.0293 & 0.0187 & 0.0460 & 0.0241
      & 0.0450 & 0.0302 & 0.0675 & 0.0374 \\
    \midrule
    \multirow{5}{*}{LLM-based}
      & P5-CID
      & 0.0465 & 0.0325 & 0.0668 & 0.0391
      & 0.0295 & 0.0214 & 0.0420 & 0.0254
      & 0.0223 & 0.0143 & 0.0357 & 0.0186 \\
      & LC-Rec
      & 0.0503 & 0.0352 & 0.0715 & 0.0420
      & 0.0259 & 0.0175 & 0.0384 & 0.0216
      & 0.0543 & 0.0385 & 0.0753 & 0.0453 \\
      & EAGER-LLM
      & 0.0548 & 0.0369 & 0.0830 & 0.0459
      & 0.0373 & 0.0251 & \underline{0.0569} & 0.0315
      & 0.0510 & 0.0349 & 0.0772 & 0.0433 \\
      & GRAM
      & 0.0641 & 0.0451 & 0.0890 & 0.0531
      & 0.0375 & 0.0256 & 0.0554 & 0.0314
      & 0.0718 & 0.0516 & 0.0987 & 0.0603 \\
      & \cellcolor{gray!12}\textbf{\method}
      & \cellcolor{gray!12}\textbf{0.0729} & \cellcolor{gray!12}\textbf{0.0508}
      & \cellcolor{gray!12}\textbf{0.1004} & \cellcolor{gray!12}\textbf{0.0597}
      & \cellcolor{gray!12}\textbf{0.0436} & \cellcolor{gray!12}\textbf{0.0301}
      & \cellcolor{gray!12}\textbf{0.0608} & \cellcolor{gray!12}\textbf{0.0356}
      & \cellcolor{gray!12}\textbf{0.0792} & \cellcolor{gray!12}\textbf{0.0578}
      & \cellcolor{gray!12}\textbf{0.1047} & \cellcolor{gray!12}\textbf{0.0660} \\
    \cmidrule(l){2-14}
      & \textit{Improv.}
      & 7.19\% & 2.48\%
      & 7.76\% & 3.46\%
      & 16.19\% & 11.92\%
      & 6.84\% & 10.65\%
      & 6.28\% & 4.35\%
      & 5.95\% & 4.28\% \\
    \bottomrule
  \end{tabularx}
\end{table*}

\paragraph{Baselines.}
We compare three families: ID-based sequential models, SID-based generative models, and LLM-based recommenders.
\begin{itemize}
  \item \textbf{SASRec}~\citep{kang2018sasrec} uses causal self-attention over item IDs to predict the next item. We train it with binary cross-entropy and one sampled negative per positive.
  \item \textbf{SASRec-CE}~\citep{klenitskiy2023dross} retains the SASRec architecture but uses cross-entropy over all item IDs. We also use SASRec-CE as M\&B's frozen teacher to propose synthetic target items during training.
  \item \textbf{BERT4Rec}~\citep{sun2019bert4rec} learns bidirectional representations of interaction sequences by predicting randomly masked items.
  \item \textbf{TIGER}~\citep{rajput2023tiger} quantizes Sentence-T5 item embeddings into hierarchical SIDs and trains a T5-style encoder--decoder to generate the next item's SID.
  \item \textbf{LETTER}~\citep{wang2024letter} learns item identifiers from semantics and SASRec collaborative embeddings, using a diversity objective to reduce code assignment bias. We use the LETTER variant built on TIGER's T5 architecture.
  \item \textbf{EAGER}~\citep{wang2024eager} shares an encoder between decoders for behavioral and semantic identifiers. It uses DIN behavioral embeddings and combines both decoders' predictions.
  \item \textbf{ETEGRec}~\citep{liu2025etegrec} tokenizes SASRec item embeddings and alternates optimization of the tokenizer and generative recommender, aligning item codes with user preferences.
  \item \textbf{P5-CID}~\citep{hua2023index} uses collaborative item identifiers in T5-based P5 to formulate recommendation as text-to-text generation.
  \item \textbf{LC-Rec}~\citep{zheng2024lcrec} aligns learned item indices and collaborative signals with an LLM for generation over the full catalog.
  \pagebreak[0]
  \begin{samepage}
  \item \textbf{EAGER-LLM}~\citep{hong2025eagerllm} incorporates DIN behavioral embeddings and semantic information into item indices and aligns these signals with a decoder-only LLM.
  \item \textbf{GRAM}~\citep{lee2025gram} adds SASRec information to item descriptions and fuses separately encoded prompts during decoding.
  \end{samepage}
\end{itemize}
Results come from our runs, the original papers, or published reproductions. Appendix~\ref{sec:appendix} records the sources and settings.

\begin{figure}[t]
  \centering
  \includegraphics[width=\columnwidth]{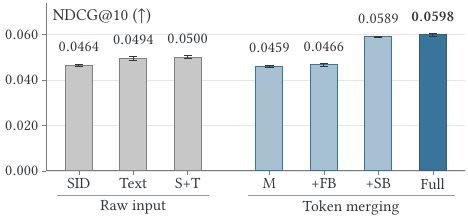}
  \caption{Component ablations on Beauty ($N_q=4$). S+T: SID+Text; M: token merging; +FB/+SB: future/synthetic supervision; Full: both. Error bars: SD over five seeds.}
  \label{fig:component-ablation}
  \Description{Bar chart of NDCG at rank ten for SID, text, and combined inputs, followed by token merging and future or synthetic supervision. Error bars show variation across runs.}
\end{figure}

\begin{figure}[t]
  \centering
  \captionsetup[subfigure]{font=footnotesize,labelfont=bf,skip=3pt,justification=centering}
  \begin{subfigure}[t]{0.488\linewidth}
    \centering
    \includegraphics[page=1,width=\linewidth]{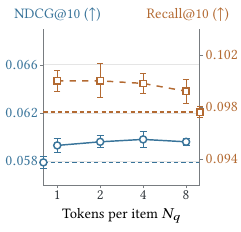}
    \caption{Accuracy}
  \end{subfigure}\hfill
  \begin{subfigure}[t]{0.488\linewidth}
    \centering
    \includegraphics[page=2,width=\linewidth]{figures/ablation_query_efficiency_single.pdf}
    \caption{Training efficiency}
  \end{subfigure}
  \caption{Token-budget ablation on Beauty. Reference lines: uncompressed SID+Text (smaller micro-batch, same effective batch). Error bars: SD over five seeds.}
  \label{fig:merge-ablation}
  \Description{Two panels compare recommendation accuracy and training cost as the number of merged tokens per item increases. The accuracy panel shows NDCG and Recall at rank ten; the efficiency panel shows optimizer-step time and peak memory. Dashed horizontal lines mark the uncompressed reference.}
\end{figure}

\paragraph{Results.}
We compare M\&B with sequential and generative recommenders on three Amazon domains (Table~\ref{tab:public-results}). M\&B consistently outperforms all baselines across all four metrics while compressing each item's SID and text into a single token. Notably, SASRec-CE provides a particularly strong baseline, matching or exceeding the other SID-based and LLM-based methods. Across all three domains and four metrics, M\&B improves over the strongest baseline for each metric by 2.48\%–16.19\%. The following ablations examine in depth how input representation and multi-target supervision contribute to this performance.

\subsection{Ablations (RQ2)}

\paragraph{Components.}
We compare SID, text, and combined inputs, then add token merging and future and synthetic supervision to examine their contributions (Figure~\ref{fig:component-ablation}; Table~\ref{tab:component-ablation}). Combining SID and text is more effective than either input alone under next-item supervision, but compression alone does not retain this advantage. Within the compact model, synthetic targets provide the largest gain, and adding future targets further improves recommendation accuracy.
\paragraph{Token merging.}
We vary the number of merged tokens per item, $N_q\in\{1,2,4,8\}$, while retaining all supervision sources (Figure~\ref{fig:merge-ablation}; Table~\ref{tab:merge-ablation}). Increasing $N_q$ raises sequence length and training cost without consistently improving accuracy. We use $N_q=1$ in the main benchmark to balance accuracy and efficiency.

\begin{figure}[t]
  \centering
  \includegraphics[width=0.9\columnwidth]{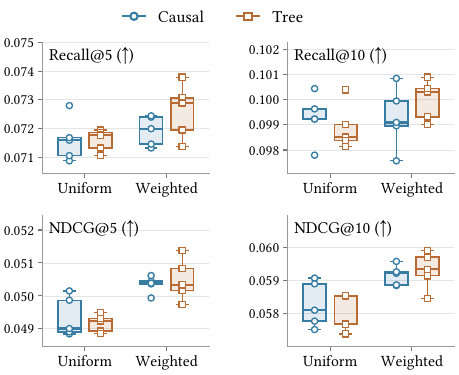}
  \caption{Target layout and loss weighting on Beauty ($N_q=1$). Each marker represents one of five seeds; Uniform/Weighted denote uniform/group weighting.}
  \label{fig:layout-ablation}
  \Description{Four panels compare Recall and NDCG for causal and tree target layouts under uniform and group-weighted losses. Box plots summarize the runs, and individual markers show each run.}
\end{figure}

\paragraph{Target layout.}
We compare causal and tree layouts under uniform and group weighting to separate the effects of target dependencies and loss weights (Figure~\ref{fig:layout-ablation}). Uniform weighting assigns equal weight to all targets. Group weighting uses next-item weight 1 and future/synthetic group weights $\lambda_{\mathrm F}=0.25$ and $\lambda_{\mathrm S}=1.0$, shared equally among valid targets within each group.
Group weighting improves NDCG for both layouts, and the weighted tree layout achieves the highest mean scores. Thus, additional targets improve recommendation without requiring dependencies between them, while balancing their contributions matters for either layout.

\subsection{Efficiency (RQ3)}

\paragraph{End-to-end training.}
We measure training step time and peak memory for merged and uncompressed SID+Text histories on H200, using an effective batch size of 128 (Figure~\ref{fig:merge-ablation}b). The uncompressed reference uses gradient accumulation. We average a stable 20-step window after warm-up for each of five seeds, excluding compilation. Compact histories reduce both costs, and adding auxiliary branches costs much less than processing the uncompressed history (Table~\ref{tab:component-ablation}). Measurement details are in Appendix~\ref{sec:attention-implementation}.

\begin{figure}[t]
  \centering
  \includegraphics[width=\linewidth]{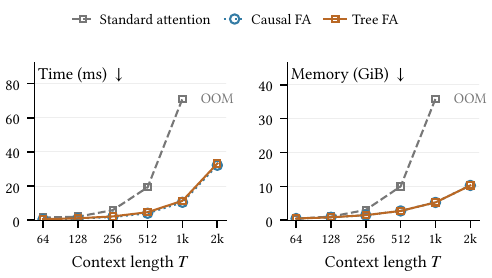}
  \caption{Attention forward--backward time and memory on H200 ($B=128$). Means over five repeats; memory: peak allocated; OOM: out of memory. Causal FA uses a causal mask rather than a tree mask.}
  \label{fig:tree-fa-scaling}
  \Description{Two line charts compare attention forward-backward time and peak memory as context length increases. Standard attention, Causal FA, and Tree FA use distinct line and hollow-marker styles. The longest standard-attention case is marked out of memory.}
\end{figure}

\paragraph{Attention implementation.}
We compare the runtime and peak memory of attention forward and backward passes across context lengths, using four independent branches of four tokens and batch size 128 (Figure~\ref{fig:tree-fa-scaling}). Standard attention uses the same tree mask; Causal FA provides a performance reference with a causal mask. Tree FA supports independent branches with modest runtime overhead and the same peak allocated memory as Causal FA. Avoiding a dense attention matrix also makes tree FA faster and more memory-efficient than standard attention at long contexts. Tree FA runs at the longest context, where standard attention runs out of memory.

\begin{table}[t]
  \centering
  \setlength{\tabcolsep}{1.5pt}
  \renewcommand{\arraystretch}{1.06}
  \caption{SID beam search on H200 ($B=1$, 100 beams, context length $T$). Causal: causal FA with duplicated per-beam KV caches; Tree: tree FA with a shared-prefix KV cache. Mean $\pm$ SD over five repeats; memory: total peak allocated. \textit{Improv.}: reduction relative to Causal.}
  \label{tab:beam-efficiency}
  \begin{tabular*}{\linewidth}{@{\extracolsep{\fill}}rrrrrrr@{}}
    \toprule
    & \multicolumn{3}{c}{Latency (ms/request) $\downarrow$} & \multicolumn{3}{c}{Memory (GiB) $\downarrow$} \\
    \cmidrule(lr){2-4}\cmidrule(l){5-7}
    $T$ & Causal & Tree & \textit{Improv.} & Causal & Tree & \textit{Improv.} \\
    \midrule
    64 & \result{86.04}{0.48} & \result{77.12}{1.55} & 10.4\% & \result{4.02}{0.00} & \result{3.31}{0.00} & 17.6\% \\
    128 & \result{86.41}{0.71} & \result{76.33}{0.62} & 11.7\% & \result{4.70}{0.00} & \result{3.32}{0.00} & 29.4\% \\
    256 & \result{86.33}{0.44} & \result{77.06}{0.41} & 10.7\% & \result{6.07}{0.00} & \result{3.34}{0.00} & 45.0\% \\
    512 & \result{87.31}{0.99} & \result{78.04}{1.42} & 10.6\% & \result{8.86}{0.00} & \result{3.39}{0.00} & 61.8\% \\
    1024 & \result{91.13}{0.39} & \result{76.44}{0.33} & 16.1\% & \result{14.43}{0.00} & \result{3.46}{0.00} & 76.0\% \\
    2048 & \result{139.85}{0.27} & \result{83.85}{0.90} & 40.0\% & \result{25.56}{0.00} & \result{3.62}{0.00} & 85.8\% \\
    \bottomrule
  \end{tabular*}
\end{table}

\begin{figure*}[t]
  \centering
  \begin{minipage}{\textwidth}
  \includegraphics[width=\linewidth]{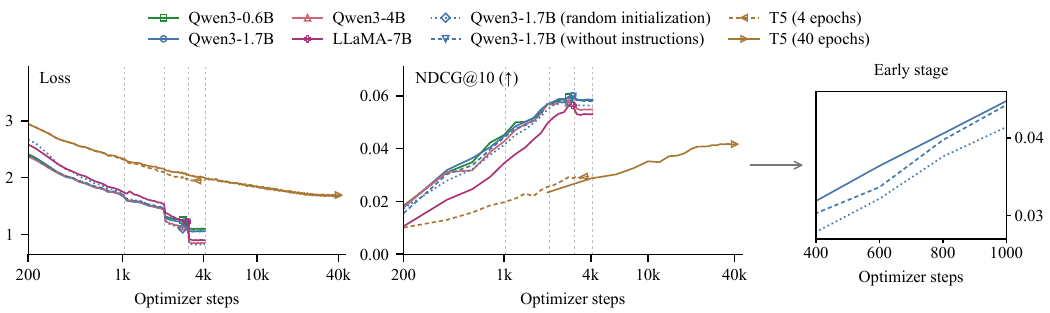}
  \par
  \captionsetup[subfigure]{font=footnotesize,labelfont=bf,skip=2pt,justification=centering}
  \makebox[\linewidth][l]{%
    \begin{subfigure}{0.37\linewidth}
      \caption{Training loss}
    \end{subfigure}%
    \begin{subfigure}{0.63\linewidth}
      \caption{Test performance}
    \end{subfigure}%
  }
  \end{minipage}
  \caption{Training curves on Beauty with different backbones. Vertical dashed lines mark the boundaries of epochs 1--4; markers denote validation-selected checkpoints.}
  \label{fig:backbone-curves}
  \Description{Training loss and test NDCG are plotted against optimizer steps for Qwen3, LLaMA, and T5 backbones and pretraining or instruction ablations. An enlarged early-stage panel compares the Qwen3-1.7B variants. Vertical dashed lines mark epoch boundaries, and hollow markers identify validation-selected checkpoints.}
\end{figure*}

\paragraph{Constrained decoding.}
We measure beam-search latency and peak memory across context lengths at beam width 100 (Table~\ref{tab:beam-efficiency}). We construct synthetic prefixes by resampling prepared history embeddings to the desired length, retaining the catalog constraints defined by the real trie. Both methods use the same checkpoint, prefixes, and decoding rules; the causal implementation copies prefix KV states for each beam, whereas our implementation stores the prefix KV states only once. Timing includes model prefill, all SID decoding steps, and beam updates; peak memory includes the model and cache. Memory savings grow with context length, and the longest contexts also show the largest speedup. The smaller speedup at short contexts is consistent with other model and search operations accounting for more of the runtime. Sharing the prefix cache is therefore especially useful for long histories. Appendix~\ref{sec:attention-implementation} describes the implementation and differences in outputs.

\subsection{Convergence (RQ4)}

\paragraph{Backbones and pretraining.}
We compare backbones through their learning curves and remove pretraining or instructions from Qwen3-1.7B (Figure~\ref{fig:backbone-curves}). Modern decoder-only LLMs, Qwen3 and LLaMA, converge faster than T5 in these experiments, reaching higher recommendation quality within fewer epochs. The TIGER-style T5 backbone continues to improve with a longer schedule, so a short training budget can understate its performance. Larger models do not always perform better: Qwen3-1.7B achieves the best test performance at selected checkpoints.

Qwen3 and LLaMA overfit late in training: training loss continues to fall while recommendation performance declines. Pretraining and instructions mainly accelerate early progress; removing either slows initial learning, but the gap narrows with further training. These trends favor selecting checkpoints by validation quality rather than training loss or a fixed epoch count.

\begin{figure}[t]
  \centering
  \includegraphics[width=\linewidth]{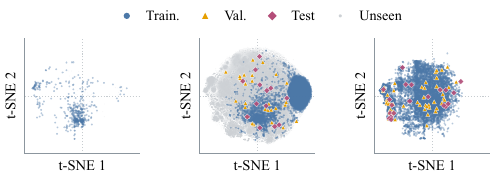}
  \par
  \captionsetup[subfigure]{font=footnotesize,labelfont=bf,skip=2pt,justification=centering}
  \begin{subfigure}[t]{0.32\linewidth}
    \caption{SID tokens}
  \end{subfigure}\hfill
  \begin{subfigure}[t]{0.32\linewidth}
    \caption{Text tokens}
  \end{subfigure}\hfill
  \begin{subfigure}[t]{0.32\linewidth}
    \caption{Merged tokens}
  \end{subfigure}
  \caption{Embedding distributions on Beauty. SID and merged embeddings are projected into the t-SNE space fitted to text embeddings. Colors indicate the data split where items first appear; Unseen denotes tokens absent from the data.}
  \label{fig:embedding-distributions}
  \Description{Three scatter plots show SID, text, and merged representations in a common t-SNE coordinate system. Colors and markers distinguish training, validation, test, and unseen tokens; merged item representations from different data splits overlap.}
\end{figure}

To examine how the merger represents items beyond the training set, we visualize SID, text, and merged representations in a common t-SNE map fitted to text embeddings (Figure~\ref{fig:embedding-distributions}). Merged representations of items first seen in validation or test mix with training items rather than forming separate clusters by data split.

\FloatBarrier

\section{Conclusion}
\label{sec:conclusion}

\method makes LLM-based recommendation more efficient by reducing the cost of text-rich histories and learning from multiple targets. Public benchmarks demonstrate strong recommendation performance alongside substantial reductions in training and decoding cost. The ablations clarify an important finding: additional training targets improve recommendation, whereas allocating more tokens to each item does not consistently help. Sharing history computation makes this richer supervision possible at a small additional cost. These results show the value of learning more from each history, rather than simply increasing the amount of input processed by the model.

The current model assigns the same token budget to every item and predicts targets independently. Future work could allocate tokens according to the information each item provides. Jointly generating recommendation lists could also capture relationships among items and balance relevance with diversity.

\bibliographystyle{ACM-Reference-Format}
\bibliography{references}

\clearpage
\appendix
\setlength{\dblfloatsep}{8pt plus 2pt minus 2pt}
\setlength{\dbltextfloatsep}{10pt plus 2pt minus 2pt}
\setcounter{dbltopnumber}{5}
\renewcommand{\dbltopfraction}{0.95}
\makeatletter
\setlength{\@dblfptop}{0pt}
\setlength{\@dblfpsep}{8pt plus 1pt minus 1pt}
\makeatother
\section{Tree FlashAttention Derivation}
\label{sec:method-details}

The following equations describe one attention head; in grouped-query attention, the key/value gradients are also summed over the query heads that share them. For a query row, let $s_n^{(t)}$ be the scaled query--key score in tile $t$ and $\mathbf v_n^{(t)}$ its value. Invisible positions have score $-\infty$. Starting with $m^{(0)}=-\infty$, $z^{(0)}=0$, and $\mathbf a^{(0)}=\mathbf 0$, the online softmax recurrence is
\begin{equation}
  \begin{aligned}
  m^{(t)}
    &=\max\!\left(m^{(t-1)},\max_n s_n^{(t)}\right), \\
  \rho^{(t)}
    &=\exp\!\left(m^{(t-1)}-m^{(t)}\right), \\
  w_n^{(t)}
    &=\exp\!\left(s_n^{(t)}-m^{(t)}\right), \\
  z^{(t)}
    &=\rho^{(t)}z^{(t-1)}+\sum_n w_n^{(t)}, \\
  \mathbf a^{(t)}
    &=\rho^{(t)}\mathbf a^{(t-1)}
      +\sum_n w_n^{(t)}\mathbf v_n^{(t)}.
  \end{aligned}
\end{equation}
Tiles with no visible entries for the row leave its state unchanged. After the last tile $J$, the output is $\mathbf o=\mathbf a^{(J)}/z^{(J)}$, and the saved log-sum-exp is $\zeta=m^{(J)}+\log z^{(J)}$. Joint normalization across prefix and branch tiles gives a single softmax over the visible keys.

The backward pass recomputes the visible probabilities as $A_{ab}=\exp(s_{ab}-\zeta_a)$. Let $\mathbf g_a=\nabla_{\mathbf o_a}\loss$ and define the score derivative
\begin{equation}
  D_{ab}=A_{ab}\,\mathbf g_a^{\top}(\mathbf v_b-\mathbf o_a),
  \label{eq:tree-score-gradient}
\end{equation}
with $D_{ab}=0$ outside the visibility set. At head dimension $d_h$, blockwise accumulation gives
\begin{equation}
  \begin{aligned}
  \nabla_{\mathbf Q_p}\loss
    &=\frac{1}{\sqrt{d_h}}\,\mathbf D_{pp}\mathbf K_p, \\[2pt]
  \nabla_{\mathbf Q_j}\loss
    &=\frac{1}{\sqrt{d_h}}\bigl(\mathbf D_{jp}\mathbf K_p+
             \mathbf D_{jj}\mathbf K_j\bigr), \\[2pt]
  \nabla_{\mathbf K_p}\loss
    &=\frac{1}{\sqrt{d_h}}\Bigl(
       \mathbf D_{pp}^{\top}\mathbf Q_p+
       \sum_j\mathbf D_{jp}^{\top}\mathbf Q_j\Bigr), \\[2pt]
  \nabla_{\mathbf V_p}\loss
    &=\mathbf A_{pp}^{\top}\mathbf G_p+
       \sum_j\mathbf A_{jp}^{\top}\mathbf G_j.
  \end{aligned}
  \label{eq:tree-block-gradients}
\end{equation}
Branch keys and values receive contributions only from their own branch queries. The gradient matrix $\mathbf G$ stacks rows $\mathbf g_a^{\top}$ and already includes the normalized supervision weights. These formulas express the computation in blocks; the kernel reconstructs and consumes tile probabilities without storing the full $\mathbf A$ or $\mathbf D$.

With matching positions and zero dropout, the packed graph has the same visible predecessors as independently executed branches. At each layer, its single prefix represents the identical prefix computation in those separate sequences; the accumulated adjoints collect every use of that computation. The graphs thus give the same weighted objective and gradients in exact arithmetic, but floating-point outputs need not be identical.

\section{Experimental Details}
\label{sec:appendix}

\subsection{Baselines}

SASRec, SASRec-CE, BERT4Rec, ETEGRec, LETTER on Sports and Toys, EAGER-LLM on Toys, and M\&B are our runs. TIGER, LETTER on Beauty, EAGER, EAGER-LLM on Beauty and Sports, and GRAM are taken from their original papers. P5-CID and LC-Rec are taken from the published GRAM reproduction because their original evaluations do not cover the same three Amazon domains.

We reproduce LETTER on Sports and Toys with the official LETTER-TIGER implementation, ETEGRec on all three datasets with its alternating tokenizer--recommender procedure, and EAGER-LLM on Toys with its released implementation. We follow the recommended settings for each implementation.

SASRec uses binary cross-entropy with one uniformly sampled negative per positive item. SASRec-CE uses cross-entropy over the full item catalog, following prior work~\citep{klenitskiy2023dross}. BERT4Rec uses masked-item prediction. All three models use Adam. We search history length in $\{20,50\}$, hidden size/head count in $\{32/1,\allowbreak64/2,\allowbreak128/4,\allowbreak256/8\}$, layers in $\{2,4\}$, learning rate in $\{10^{-4},3\times10^{-4},10^{-3}\}$, weight decay in $\{0,10^{-2}\}$, and dropout in $\{0.2,0.5\}$. BERT4Rec additionally searches mask ratio in $\{0.2,0.4,0.6\}$. For each dataset, we select the run with the highest validation NDCG@10 and report its test metrics.

\begin{table*}[t]
  \centering
  \setlength{\tabcolsep}{4pt}
  \caption{Attention forward--backward cost on H200 ($B=128$). Mean $\pm$ SD over five repeats; memory: peak allocated; OOM: out of memory. Causal FA uses a causal mask.}
  \label{tab:tree-fa-scaling}
  \begin{tabular}{@{}lrrrrrr@{}}
    \toprule
    Context length & \multicolumn{3}{c}{Time (ms)} & \multicolumn{3}{c}{Memory (GiB)} \\
    \cmidrule(lr){2-4}\cmidrule(l){5-7}
    & Standard attention & Tree FA & Causal FA & Standard attention & Tree FA & Causal FA \\
    \midrule
    64 & 1.564 $\pm$ 0.042 & 0.590 $\pm$ 0.001 & 0.545 $\pm$ 0.003 & 0.475 $\pm$ 0.000 & 0.502 $\pm$ 0.000 & 0.502 $\pm$ 0.000 \\
    128 & 2.112 $\pm$ 0.001 & 1.160 $\pm$ 0.002 & 0.980 $\pm$ 0.002 & 1.082 $\pm$ 0.000 & 0.878 $\pm$ 0.000 & 0.878 $\pm$ 0.000 \\
    256 & 5.925 $\pm$ 0.001 & 2.237 $\pm$ 0.002 & 1.872 $\pm$ 0.001 & 3.051 $\pm$ 0.000 & 1.505 $\pm$ 0.000 & 1.505 $\pm$ 0.000 \\
    512 & 19.277 $\pm$ 0.000 & 4.712 $\pm$ 0.003 & 4.080 $\pm$ 0.002 & 9.989 $\pm$ 0.000 & 2.759 $\pm$ 0.000 & 2.759 $\pm$ 0.000 \\
    1,024 & 71.106 $\pm$ 0.015 & 11.404 $\pm$ 0.132 & 10.513 $\pm$ 0.182 & 35.865 $\pm$ 0.000 & 5.268 $\pm$ 0.000 & 5.268 $\pm$ 0.000 \\
    2,048 & OOM & 33.205 $\pm$ 0.047 & 32.183 $\pm$ 0.065 & OOM & 10.286 $\pm$ 0.000 & 10.286 $\pm$ 0.000 \\
    \bottomrule
  \end{tabular}
\end{table*}

\subsection{M\&B on Public Data}
\label{sec:public-configuration}

We fine-tune Qwen3 1.7B and use the frozen SASRec-CE checkpoint with the highest validation NDCG@10 to construct synthetic branches. Each historical item is represented by its SID, title, category, and description. The runs reported in the main benchmark use $N_q=1$. SIDs are obtained from Sentence-T5-XXL content embeddings with a 3-level, 256-way RQ-VAE and one deterministic collision-suffix token, giving SID depth $L=4$. The maximum input history length is 20. Each training window has the next observed item, at most one additional future item, and two synthetic targets. Windows with no additional future item are retained; no held-out item is used to fill a missing branch. We use group weights $\lambda_{\mathrm S}=1.0$ and $\lambda_{\mathrm F}=0.25$, distributed over the valid branches in each group as in Section~\ref{sec:multi-target-supervision}.

The token merger uses attention dimension $d_a=256$ and value dimension $d_v=512$. The MLP in the SID anchor uses a hidden dimension of 512. Public benchmarks use SID and text inputs without action-type or time-bucket embeddings.

All runs use tree attention, bfloat16, AdamW, batch size 128, weight decay $10^{-2}$, 200 warmup steps, cosine decay, 4 epochs, and beam width 50. The backbone and token merger use learning rate $10^{-4}$, while the tied token embeddings use learning rate $10^{-3}$. We evaluate the model every 500 steps and at the end of each epoch. The final checkpoint is selected by validation NDCG@10.

\subsection{Efficiency Measurement}
\label{sec:attention-implementation}

All efficiency measurements use H200. Our attention benchmark follows FA-2~\citep{dao2024flashattention2}, using BF16, 16 query heads, 8 KV heads, head dimension 128, and four branches of four tokens each. Standard attention materializes the score matrix and applies the same tree mask as tree FA. Causal FA uses FA-2 with a causal mask at matching tensor shapes and total length $T+16$. At batch size 128, we average 30 forward--backward iterations after warm-up in each of five repeats. Table~\ref{tab:tree-fa-scaling} shows that tree FA uses the same peak allocated memory as causal FA, with negligible runtime overhead across the tested contexts. In contrast, the cost of standard attention with an explicit mask grows rapidly with context length.

For decoding, we use batch size one, beam width 100, and a Qwen3-1.7B model trained for four epochs with $N_q=4$. Synthetic contexts of 64--2048 tokens are sampled from Beauty history embeddings; both methods use identical inputs and catalog constraints. CUDA-synchronized timing includes prefill and beam search, excluding shared input preparation and trie construction. Memory includes the model, inputs, cache, outputs, and workspace. We report mean and sample standard deviation over five repeats, averaging latency over eight prefixes per repeat. Table~\ref{tab:beam-efficiency} measures beam-search efficiency. Numerical differences may change beam candidates; the timing comparison does not assume identical output lists.

For end-to-end training in Tables~\ref{tab:component-ablation} and~\ref{tab:merge-ablation}, we use tree FA, BF16, and a compiled backbone. Each configuration uses five independent runs and effective batch size 128. We average 20 optimizer steps after warm-up, excluding compilation and retaining the same learning-rate schedule. Memory is peak allocated memory in decimal GB. We report the mean and sample standard deviation across the five runs. The component and token-budget ablation results show that auxiliary supervision improves the recommendation performance, whereas additional queries increase training cost without consistently improving accuracy.

\begin{table*}[t]
  \centering
  \setlength{\tabcolsep}{5pt}
  \renewcommand{\arraystretch}{1.0}
  \caption{Component ablations on Beauty ($N_q=4$); notation follows Figure~\ref{fig:component-ablation}. Mean $\pm$ SD over five runs; bold: best accuracy. Time: optimizer-step latency; memory: peak allocated. $^{*}$: micro-batch 16, accumulation 8.}
  \label{tab:component-ablation}
  \begin{tabular}{@{}lcccrrrrrr@{}}
    \toprule
    Input & M & FB & SB & Recall@5 & NDCG@5 & Recall@10 & NDCG@10 & Time (ms) & Memory (GB) \\
    \midrule
    SID & & & & \result{0.0555}{0.0009} & \result{0.0390}{0.0006} & \result{0.0786}{0.0011} & \result{0.0464}{0.0006} & \result{279.92}{10.07} & \result{42.96}{0.00} \\
    Text & & & & \result{0.0584}{0.0015} & \result{0.0413}{0.0008} & \result{0.0837}{0.0015} & \result{0.0494}{0.0009} & \result{3611.06}{71.56}$^{*}$ & \result{99.19}{0.00}$^{*}$ \\
    SID + Text & & & & \result{0.0594}{0.0006} & \result{0.0421}{0.0005} & \result{0.0842}{0.0014} & \result{0.0500}{0.0006} & \result{3674.85}{72.76}$^{*}$ & \result{99.19}{0.00}$^{*}$ \\
    SID + Text & $\checkmark$ & & & \result{0.0546}{0.0011} & \result{0.0375}{0.0007} & \result{0.0808}{0.0005} & \result{0.0459}{0.0005} & \result{289.57}{1.02} & \result{44.16}{0.04} \\
    SID + Text & $\checkmark$ & $\checkmark$ & & \result{0.0554}{0.0010} & \result{0.0381}{0.0008} & \result{0.0819}{0.0014} & \result{0.0466}{0.0007} & \result{295.12}{1.08} & \result{45.19}{0.04} \\
    SID + Text & $\checkmark$ & & $\checkmark$ & \result{0.0719}{0.0008} & \result{0.0505}{0.0006} & \result{0.0982}{0.0005} & \result{0.0589}{0.0004} & \result{313.78}{1.67} & \result{47.27}{0.04} \\
    SID + Text & $\checkmark$ & $\checkmark$ & $\checkmark$ & \result{\textbf{0.0729}}{0.0005} & \result{\textbf{0.0511}}{0.0006} & \result{\textbf{0.0998}}{0.0008} & \result{\textbf{0.0598}}{0.0007} & \result{340.68}{36.33} & \result{47.27}{0.04} \\
    \bottomrule
  \end{tabular}
\end{table*}

\begin{table*}[t]
  \centering
  \setlength{\tabcolsep}{5pt}
  \renewcommand{\arraystretch}{1.0}
  \caption{Token-budget ablation on Beauty with full supervision. Len.: mean input tokens. Conventions and $^{*}$ follow Table~\ref{tab:component-ablation}.}
  \label{tab:merge-ablation}
  \begin{tabular}{@{}lrrrrrrr@{}}
    \toprule
    Compression & Len. & Recall@5 & NDCG@5 & Recall@10 & NDCG@10 & Time (ms) & Memory (GB) \\
    \midrule
    Raw input & 666.6 & \result{0.0711}{0.0002} & \result{0.0494}{0.0005} & \result{0.0976}{0.0004} & \result{0.0579}{0.0005} & \result{3650.01}{55.51}$^{*}$ & \result{99.58}{0.00}$^{*}$ \\
    Mean pooling & 31.3 & \result{0.0711}{0.0002} & \result{0.0495}{0.0002} & \result{0.0970}{0.0007} & \result{0.0578}{0.0002} & \result{185.66}{2.55} & \result{30.58}{0.00} \\
    Merger, $N_q=1$ & 31.3 & \result{0.0726}{0.0010} & \result{0.0505}{0.0006} & \result{\textbf{0.1000}}{0.0008} & \result{0.0593}{0.0006} & \result{192.34}{2.57} & \result{31.59}{0.04} \\
    Merger, $N_q=2$ & 38.5 & \result{0.0728}{0.0008} & \result{0.0509}{0.0004} & \result{\textbf{0.1000}}{0.0013} & \result{0.0596}{0.0005} & \result{227.53}{2.95} & \result{36.82}{0.04} \\
    Merger, $N_q=4$ & 52.8 & \result{\textbf{0.0729}}{0.0005} & \result{\textbf{0.0511}}{0.0006} & \result{0.0998}{0.0008} & \result{\textbf{0.0598}}{0.0007} & \result{340.68}{36.33} & \result{47.27}{0.04} \\
    Merger, $N_q=8$ & 81.4 & \result{0.0724}{0.0003} & \result{0.0509}{0.0004} & \result{0.0992}{0.0009} & \result{0.0596}{0.0003} & \result{464.41}{30.91} & \result{68.19}{0.05} \\
    \bottomrule
  \end{tabular}
\end{table*}

\begin{table*}[t]
  \centering
  \setlength{\tabcolsep}{7pt}
  \renewcommand{\arraystretch}{1.0}
  \caption{Target layout and loss weighting on Beauty ($N_q=1$). Mean $\pm$ SD over five runs; bold: best results.}
  \label{tab:layout-ablation-full}
  \begin{tabular}{@{}llrrrr@{}}
    \toprule
    Layout & Loss & Recall@5 & NDCG@5 & Recall@10 & NDCG@10 \\
    \midrule
    Causal & Uniform & \result{0.0716}{0.0008} & \result{0.0494}{0.0006} & \result{0.0993}{0.0010} & \result{0.0583}{0.0007} \\
    Tree & Uniform & \result{0.0716}{0.0004} & \result{0.0492}{0.0003} & \result{0.0989}{0.0009} & \result{0.0580}{0.0005} \\
    Causal & Weighted & \result{0.0719}{0.0005} & \result{0.0504}{0.0003} & \result{0.0993}{0.0012} & \result{0.0592}{0.0003} \\
    Tree & Weighted & \result{\textbf{0.0726}}{0.0010} & \result{\textbf{0.0505}}{0.0006} & \result{\textbf{0.1000}}{0.0008} & \result{\textbf{0.0593}}{0.0006} \\
    \bottomrule
  \end{tabular}
\end{table*}

\subsection{Auxiliary Branch Counts and Weights}

Table~\ref{tab:auxiliary-weights} studies future and synthetic supervision separately, each added to the next-item objective. FB$\times1$ uses at most one available future item. SB$\times k$ uses $k$ synthetic branches sharing group weight $\lambda_{\mathrm S}$, with future supervision disabled. A lighter future loss is more effective, while synthetic supervision benefits from a moderate weight: the best tested settings are $\lambda_{\mathrm F}=0.25$ and SB$\times2$ with $\lambda_{\mathrm S}=1.0$. Adding more synthetic branches at the same total weight improves Recall@10 but reduces NDCG@10, suggesting a trade-off between retrieval coverage and top-ranked relevance.

\begin{table*}[t]
  \centering
  \setlength{\tabcolsep}{7pt}
  \renewcommand{\arraystretch}{1.0}
  \caption{Number and weights of auxiliary branches on Beauty ($N_q=1$). Mean $\pm$ SD over five runs. FB/SB: future/synthetic branches; shading: selected settings; bold: best results.}
  \label{tab:auxiliary-weights}
  \begin{tabular}{@{}lrrrr@{}}
\toprule
Setting & Recall@5 & NDCG@5 & Recall@10 & NDCG@10 \\
\midrule
Next-item only & \result{0.0497}{0.0020} & \result{0.0337}{0.0018} & \result{0.0747}{0.0027} & \result{0.0417}{0.0021} \\
\midrule
\rowcolor{gray!12}
+FB$\times1$, $\lambda_{\mathrm F}=0.25$ & \result{0.0527}{0.0020} & \result{0.0355}{0.0014} & \result{0.0786}{0.0026} & \result{0.0439}{0.0016} \\
+FB$\times1$, $\lambda_{\mathrm F}=0.5$ & \result{0.0517}{0.0028} & \result{0.0348}{0.0018} & \result{0.0776}{0.0029} & \result{0.0432}{0.0018} \\
+FB$\times1$, $\lambda_{\mathrm F}=1.0$ & \result{0.0485}{0.0016} & \result{0.0325}{0.0013} & \result{0.0732}{0.0016} & \result{0.0404}{0.0013} \\
\midrule
+SB$\times2$, $\lambda_{\mathrm S}=0.5$ & \result{0.0703}{0.0009} & \result{0.0495}{0.0007} & \result{0.0976}{0.0012} & \result{0.0583}{0.0006} \\
\rowcolor{gray!12}
+SB$\times2$, $\lambda_{\mathrm S}=1.0$ & \result{0.0725}{0.0006} & \result{\textbf{0.0510}}{0.0006} & \result{0.0981}{0.0016} & \result{0.0592}{0.0008} \\
+SB$\times2$, $\lambda_{\mathrm S}=2.0$ & \result{0.0714}{0.0006} & \result{0.0493}{0.0002} & \result{0.0969}{0.0015} & \result{0.0576}{0.0004} \\
+SB$\times4$, $\lambda_{\mathrm S}=2.0$ & \result{0.0718}{0.0009} & \result{0.0482}{0.0008} & \result{0.0994}{0.0010} & \result{0.0571}{0.0008} \\
\midrule
Full M\&B & \result{\textbf{0.0726}}{0.0010} & \result{0.0505}{0.0006} & \result{\textbf{0.1000}}{0.0008} & \result{\textbf{0.0593}}{0.0006} \\
\bottomrule
\end{tabular}

\end{table*}

\end{document}